# Comparison of Markowitz Model and Single-Index Model on Portfolio Selection of Malaysian Stocks

Zhang Chern Lee[1], Wei Yun Tan[1], Hoong Khen Koo[1], Wilson Pang[1]

[1]Institute of Mathematical Sciences, Faculty of Science, Universiti Malaya, 50603 Kuala Lumpur, Malaysia

## Abstract

Our article is focused on the application of Markowitz Portfolio Theory and the Single Index Model on 10-year historical monthly return data for 10 stocks included in FTSE Bursa Malaysia KLCI, which is also our market index, as well as a risk-free asset which is the monthly fixed deposit rate. We will calculate the minimum variance portfolio and maximum Sharpe portfolio for both the Markowitz model and Single Index model subject to five different constraints, with the results presented in the form of tables and graphs such that comparisons between the different models and constraints can be made. We hope this article will help provide useful information for future investors who are interested in the Malaysian stock market and would like to construct an efficient investment portfolio.

## 1. Introduction

Bursa Malaysia, formerly known as the Kuala Lumpur Stock Exchange (KLSE) is the Malaysian stock exchange and one of the largest stock markets in Southeast Asia. In recent times, much awareness has been raised on the importance of investing. Many investors nowadays are interested in learning how to construct mathematically efficient portfolios by using various models such as the Markowitz model developed by Harry Markowitz in 1952 (Markowitz, 1952), as well as the Single Index Model developed by William F. Sharpe in 1963 (Sharpe, 1962). These models form the basis of what is now known as Modern Portfolio Theory (MPT). Using these models helps investors to reduce their portfolio risk through diversification and finding the optimal weights for each asset. Compared to traditional investing methods, reducing risk through diversification and mathematical rigour are the main advantages of MPT.

This research aims to analyse the portfolios constructed by using the Markowitz model (MM) as well as the Single Index model (IM), subject to different constraints similar to the regulations imposed on the market in the real world. The ten stocks that we have selected for our research are as follows:

| Ticker | Full Name | Sector |
|---|---|---|
| AXIA | Axiata Group Berhad | Telecommunications & Media |
| CELC | CelcomDigi Berhad | Telecommunications & Media |
| INAR | Inari Amertron Berhad | Technology |
| DIAL | Dialog Group Berhad | Energy |
| GENT | Genting Berhad | Consumer Products & Services |
| GENM | Genting Malaysia Berhad | Consumer Products & Services |
| HTHB | Hartalega Holdings Berhad | Health Care |
| HLBB | Hong Leong Bank Berhad | Financial Services |
| HLCB | Hong Leong Capital Berhad | Financial Services |
| IHHH | IHH Healthcare Berhad | Health Care |

**Table 1:** List of stocks

Our calculations use historical daily return data from 1/1/2013 to 1/8/2023 aggregated monthly to account for non-Gaussian effects. Based on the monthly return data, we prepared an Excel spreadsheet to calculate the minimum variance portfolio and maximum Sharpe portfolio of the two different models for the five different constraints. We then plot the efficient frontier and capital allocation lines for the two different models. The Excel Solver function was used to calculate the results for the models with constraints due to the large number of assets involved. The results of the two models are then compared.

This paper also aims to show the differences in the results obtained by using MM and IM. This will provide insight for aspiring investors on which model better suits their requirements, which will help them construct more efficient portfolios. It is also useful to investors who wish to invest in the Malaysian stock market, as all our stocks and market index are based on the FTSE Bursa Malaysia KLCI.

# 2. Literature Review and Methodology

**Markowitz Model (MM)**

Harry Markowitz first introduced his model of mean-variance portfolio analysis in his 1952 paper *Portfolio Selection*. The theory caused a profound scientific revolution in finance. The Markowitz model is a mathematical procedure to determine the optimum portfolios in which to invest. (Francis & Kim, 2013, p. 85)

The Markowitz Model requires the following statistical inputs:

- The expected return for each investment candidate
- The standard deviation of returns for each investment candidate
- The correlation coefficients between all pairs of investment candidates

The model takes all statistical inputs listed above and analyses them simultaneously to determine a series of plausible investment portfolios. Each solution also gives exact portfolio weightings for the investment candidates in that solution. Assumptions for the model are that all each investment opportunity is represented by a probability distribution of returns measured over the same holding period, investor risk estimates are proportional to variance of returns, investors base their decisions only on expected return and risk statistics and all investors are risk-averse rate of return maximisers.

Markowitz formulated a model grounded in four foundational assumptions to establish a robust and operational framework.

- In evaluating each investment alternative, investors analyse it by considering the probability distribution that characterises the returns on securities within a specific time horizon.
- Investors gauge the risk of a security portfolio based on the variance or standard deviation of the expected return rate of the security.
- Decisions made by investors are exclusively rooted in the assessment of risks and returns associated with the considered securities.
- At a defined risk level, investors anticipate the maximum possible return; conversely, at a specific level of return, investors exhibit a preference for minimised risk. Essentially, Markowitz's model posits that investors aim for an optimal equilibrium between risk and return, underscoring the principles of diversification and risk aversion.

Based on these four assumptions, the Markowitz Model can calculate the target outcomes of investors by the following formulas (Francis & Kim, 2013, pp. 2-3):

1. Mean:

$$E(r_i) = \sum_{s=1}^{S} p_s r_{i,s}$$

2. Risk:

$$\sigma_i^2 = E[r_i - E(r_i)]^2 = \sum_{s=1}^{S} p_s [r_s - E(r_i)]^2$$

**Single Index Model (IM)**

Harry Markowitz laid down the foundation of the Modern Portfolio Theory in 1952. Although his model was theoretically sound, it had certain limitations which were later rectified by his own protégé William Sharpe through his single-index model. The original model given by Markowitz was based on the premise that there are gains from diversification. He propounded through his theory, a method for diversification of securities. Grouping securities with negative relationships given by covariance or correlation coefficients was imperative to his optimisation method. This model required a large number of inputs. For $n$ securities, the number of inputs according to the Markowitz model would be $\frac{n^2-2}{2}$. The quality of these inputs would also affect the optimum portfolio. For example, a classic failure of the model would be when there was a portfolio with three securities A, B, and C with weights 1, 1, and −1 and each having a standard deviation of 20 percent. In this situation, the portfolio variance would be −200 — an absurd result as risk can never be a negative number. Another problem was the concept of selecting securities having negative covariances. In the real world, securities tend to move together and hence are found to have positive covariances. Since markets are often moved by sentiments, securities tend to move together in one direction. Empirical evidence also showed that portfolios with securities having positive covariances outperformed Markowitz's optimum portfolios.

Owing to the limitations discussed and the contrary empirical evidence, the need for a simpler method for portfolio selection had become imperative. William Sharpe was a doctoral student at UCLA majoring in economics and finance. When the time came for Sharpe to write his thesis, Fred Weston suggested that he should meet Markowitz. Thus, Markowitz became Sharpe's unofficial thesis advisor. Markowitz put him to work and asked him to find a simpler method for portfolio selection and optimisation. Sharpe simplified the model which we now know as the 'market model' or the single-index model. Sharpe said that instead of comparing each security with another security and trying to find negative covariances and correlations between individual securities for portfolio selection, securities should be compared to some common index. This gave birth to the concept of the market index. Sharpe reasoned that common economic factors such as business cycles, interest rates, technology changes, cost of labour, raw material, inflation, weather conditions, etc., affected the performance of all firms. Unexpected changes in these variables would cause unexpected changes in the prices and returns of all the stocks in the market. Sharpe proposed that all the economic factors could be summarised by one macroeconomic indicator which would move the entire market. Further, it was assumed that all other uncertainties in stock returns were firm-specific, i.e., there was no other form of correlation between the securities. Firm-specific events such as profits, management quality, new inventions, etc., would only affect the fortunes of individual firms and not the whole market or the broad economy in any significant way. Thus, Sharpe proposed the concept of a single market index as the surrogate for all the other individual securities in the market. Markowitz and Sharpe were awarded the Nobel Prize for their contributions to Modern Portfolio theory. (Mistry & Khatwani, 2023, p. 2)

In developing the single-index model, five assumptions were made about the random error term.

- The expected value of the random error term is zero, $E(\varepsilon_{it}) = 0$.
- The error terms are homoscedastic.
- $r_{mt}$ is not correlated with the random error term. $Cov(r_{mt}, \varepsilon_{it}) = 0$.
- The random error terms are serially uncorrelated. $Cov(\varepsilon_{it}, \varepsilon_{is}) = 0$ for $t$ not equal to $s$.
- The random error terms of an asset are not correlated with those of any other asset. $Cov(\varepsilon_{it}, \varepsilon_{jt}) = 0$ for $i$ not equal to $j$.

Based on these five assumptions, Markowitz's return-generating function can be expressed as follows:

$$r_{it} = \alpha_i + \beta_i r_{mt} + \varepsilon_{it}$$

where total return ($r_{it}$) is a combination of a systematic part ($\alpha_i + \beta_i r_{mt}$) which can be systematically explained by the market return ($r_{mt}$) and an unsystematic part ($\varepsilon_{it}$) which cannot be explained by the market return. (Francis & Kim, 2013, pp. 165-173)

On the basis of Markowitz's proposed return-generating function, Jensen (1972) proposed a similar return-generating function by changing the random variable in the original function from holding-period return ($r_{it}$) to risk premium ($r_{it} - r_f$).

$$r_{it} - r_f = \alpha'_i + \beta_i(r_{mt} - r_f) + \varepsilon_{it}$$

where $r_f$ is the risk-free rate (Jensen, 1972).

**Recent Advances and Behavioural Portfolio Theory**

The MPT is based on the basic assumptions that investors in the market choose a set of efficient mean-variance asset combinations, and they are rational, risk averse, and homogeneous. The real investors, on the contrary, have different perceptions about the market and they all are not rational at all. In the Prospect Theory, Kahneman & Tversky (1979) have found that investors who buy insurance also buy lottery tickets. Even of the dominance of the MPT during the half century of 1952-2000, the Behavioural Portfolio Theory (hereinafter BPT) has addressed a relatively new paradigm of the behavioural theories. The BPT introduces the relevance of behavioural aspects of the human beings that come within the decision-making process for portfolio selections.

Investor's psychological aspects, beliefs and preferences, changes in their portfolio choice decisions at their choice of different time frames. At the presence of the behavioural biases, the MPT has offered limited performance and the same has paved the development of the concept of behavioural portfolio theory. The said development is contemporary to the development of the theory of mental account, overconfidence, and naïve diversification. While Markowitz's Model (MM) is silent about the utility of portfolio consumption goals, these goals are central in the BPT of Shefrin & Statman (2000).

In the BPT, investors do not consider their investments in portfolio rather they consider the collection of mental account (MT) sub-portfolios. Every sub-portfolio is associated with their specific goals. In the MM, investor's asset allocation results from a trade-off between their expected returns and risk measured by the variances. For a given level of expected return, investors aim at minimising the variance of their portfolios. Investors have distinct mental accounts with different levels of aspiration. They do not consider the same as a complete portfolio rather a collection of mental accounting sub-portfolios with distinct aspiration levels. In the BPT, risk relates to the downside risk rather than the MM's indefinite forms of return variations. (Sinha & Biswas, 2018, p. 3)

In this project, we imposed distinct constraints on each model, specifically varying the upper and lower weight boundaries. These constraints mirror prevalent policies and regulations in global economic markets and companies. The incorporation of five constraints enhances the coherence and applicability of comparing and contrasting the two models.

1. $\sum|w_i| \leq 2$ :

This constraint, inspired by FINRA's Regulation T, aligns with Malaysia's need to regulate broker-dealers and ensure prudent use of customer account equity. Adapting this constraint reflects Malaysia's interest in controlling leverage and maintaining stability in its financial markets.

2. $|w_i| \leq 1$ for all i :

Reflecting client-provided "box" constraints, this ensures that individual weights remain within reasonable bounds. In the Malaysian context, this accommodates diverse investor preferences and risk tolerances, acknowledging the need for customised investment solutions.

3. “Free” problem without additional constraints:

Illustrating an unconstrained scenario allows us to understand how portfolios might behave without imposed restrictions. This insight is crucial for comprehending the potential range of investment strategies in Malaysia's dynamic and evolving financial landscape.

4. $w_i \geq 0$ for all i :

This constraint basically indicates the prohibition of short positions in the market. Although Malaysia does not fully prohibit short sales of securities, for the purpose of generalisation of this paper, we make this assumption for the purpose of generalisation of the outcomes of this paper.

5. $w_{11} = 0$ :

Investigating the impact of including a broad index in the portfolio becomes pertinent for Malaysia to assess the effects on diversification and overall performance. This constraint allows us to explore whether incorporating a broad index has positive or negative implications for Malaysian portfolios.

# 3. Result Analysis

Before diving into a more in-depth analysis, it is crucial to allocate our original data. The data collection spanned daily closing prices on trading days for 10 stocks and 1 index over the past 20 years, from January 2013 to August 2023. To make calculations simpler and ensure the data fits well with Gaussian distribution assumptions, we must initially convert the daily data into monthly observations. This conversion not only reduces computational time but also aligns with the assumption in both models that all data follow a Gaussian distribution.

Using the filter function to select the initial data point for each month and labelling them as BOM (Beginning of Month), this approach provides a representative figure for each month. Subsequently, by copying and pasting these data points into a new table, our initial data processing is completed.

For the period spanning January 2013 to August 2023, we derive the return for each stock by subtracting the current month's value from the previous month's value and then dividing by the previous month's value.

In determining the risk-free rate for our analysis, we systematically acquire the annual fixed deposit rate (1 month) for each month. Subsequently, we convert this annual rate to a monthly fixed deposit rate (1 month) to align with our monthly data framework. Utilising this monthly fixed deposit rate as our measure for the risk-free rate, we then compute the average across all months. Hence, the average monthly risk-free rate is equal to 0.002139918. This average serves as our fixed risk-free rate, serving as a consistent benchmark for subsequent calculations in our research. This meticulous process ensures a standardised and representative measure of the risk-free rate, enhancing the robustness of our analytical framework.

We randomly set a group of weights with the only constraint that the sum of weights equals one to see how two different models behave in our experiment. From the table of weights in the portfolio, we assume that the weights of the Markowitz model and Index model are identical. The weights and the results are shown in Table 2 and Table 3 as below.

**Table 2:** Weights of 10 stocks and market index

| Sum | INAR | CELC | AXIA | HLCB | HLBB | IHHH | HTHB | GENM | DIAL | GENT | Market |
|---|---|---|---|---|---|---|---|---|---|---|---|
| 1.00 | 0.04 | 0.04 | 0.04 | 0.04 | 0.04 | 0.04 | 0.04 | 0.04 | 0.04 | 0.04 | 0.60 |

**Table 3:** Results in MM & IM

| | Markowitz Model Portfolio | Index Model Portfolio |
|---|---|---|
| Return | 0.001901 | 0.001901 |
| Standard Deviation | 0.031137 | 0.031305 |
| Sharpe Ratio | -0.007686 | -0.00764 |

From Table 3, the two models do show almost identical results. Next, we need to find out the minimum variance portfolio and maximum sharpe portfolio for both models. The results are shown below in Table 4 and Table 5.

**Table 4:** Minimum variance portfolio and maximum sharpe portfolio in MM

| | INAR | CELC | AXIA | HLCB | HLBB | IHHH | HTHB | GENM | DIAL | GENT | Market | Return | StDev | Sharpe |
|---|---|---|---|---|---|---|---|---|---|---|---|---|---|---|
| Minimum Variance | 0.002192719 | 0.113491573 | -0.131888659 | -0.00372364 | 0.201990048 | 0.224155746 | -0.019076893 | 0.016076127 | -0.000807211 | -0.127410013 | 0.725000 | 0.002149317 | 0.025442836 | 0.000369427 |
| Max Sharpe | 11.43214108 | 3.411554267 | 0.803156204 | 8.487087625 | 19.78625986 | 17.83807239 | 2.199026468 | 3.991607809 | 10.55607593 | -6.635944759 | -70.86903686 | 0.74810589 | 1.746113937 | 0.427214947 |

**Table 5:** Minimum variance portfolio and maximum sharpe portfolio in IM

| | INAR | CELC | AXIA | HLCB | HLBB | IHHH | HTHB | GENM | DIAL | GENT | Market | Return | StDev | Sharpe |
|---|---|---|---|---|---|---|---|---|---|---|---|---|---|---|
| Minimum Variance | 0.002373292 | 0.052599382 | -0.103097903 | 0.05204525 | 0.155853576 | 0.192399546 | 0.004248873 | -0.038791442 | -0.03032638 | -0.072007074 | 0.784703 | 0.001731629 | 0.026158764 | -0.015608123 |
| Max Sharpe | 13.66581122 | 0.401303613 | -4.824068355 | 13.31936961 | 16.95002535 | 17.11569316 | 3.680729248 | 3.34525265 | 10.38778887 | -0.929611795 | -72.11229357 | 0.862552342 | 2.060363327 | 0.417602281 |

In employing both the Markowitz model and the Single Index model to construct minimum variance portfolios, notable differences in key performance metrics are evident. The minimum variance portfolio generated by the Markowitz model exhibits a slightly higher expected return at 0.002149317 compared to the Single Index model's return of 0.001731629. However, the Markowitz model achieves this with a lower standard deviation of 0.025442836, reflecting a potentially more efficient risk-return trade-off. Additionally, the Sharpe ratio, a measure of risk-adjusted return, for the Markowitz portfolio stands at 0.000369427, suggesting a positive risk-adjusted performance. On the other hand, the Single Index model's minimum variance portfolio, while exhibiting a marginally lower return, has a higher standard deviation of 0.026158764, resulting in a negative Sharpe ratio of -0.015608123. This implies a less favourable risk-adjusted performance compared to the Markowitz portfolio.

The maximum Sharpe portfolio derived from the Markowitz model exhibits a considerable expected return of 0.74810589, coupled with a standard deviation of 1.746113937. The resulting Sharpe ratio of 0.427214947 reflects a favourable risk-adjusted performance, suggesting a balance between return and volatility. In contrast, the maximum Sharpe portfolio generated by the Single Index model boasts a slightly higher expected return of 0.862552342 but comes with a higher standard deviation of 2.060363327. Consequently, the Sharpe ratio for the Single Index model portfolio stands at 0.417602281, indicating a slightly less efficient risk-return trade-off compared to the Markowitz portfolio.

Furthermore, we want to test whether two models can still yield identical or similar results under different additional constraints, with the default constraint being that the sum of weights equals 1.

**Constraint 1**

**Table 6**: Minimum variance portfolio and maximum sharpe portfolio in MM

| | INAR | CELC | AXIA | HLCB | HLBB | IHHH | HTHB | GENM | DIAL | GENT | Market | Return | StDev | Sharpe |
|---|---|---|---|---|---|---|---|---|---|---|---|---|---|---|
| Minimum Variance | 0.002196475 | 0.113346492 | -0.131833832 | -0.003713377 | 0.201984851 | 0.224247257 | -0.019074578 | 0.01607309 | -0.000801865 | -0.127435596 | 0.725011083 | 0.002149815 | 0.025442835 | 0.000388992 |
| Max Sharpe | 0.405124359 | 3.71515E-07 | -0.163585893 | 0.135385481 | 0.327354075 | 0.487561088 | 0.00010964 | -2.22541E-06 | 0.144464987 | -0.231511972 | -0.104899909 | 0.020470286 | 0.054598442 | 0.335730601 |

**Table 7**: Minimum variance portfolio and maximum sharpe portfolio in IM

| | INAR | CELC | AXIA | HLCB | HLBB | IHHH | HTHB | GENM | DIAL | GENT | Market | Return | StDev | Sharpe |
|---|---|---|---|---|---|---|---|---|---|---|---|---|---|---|
| Minimum | 0.00237 | 0.05261 | -0.1031 | 0.05206 | 0.15585 | 0.19240 | 0.00424 | -0.0387 | -0.0303 | -0.0720 | 0.78466 | 0.00173 | 0.02615 | -0.0156 |
| Variance | 5777 | 9921 | 03554 | 4822 | 5838 | 4374 | 7342 | 76752 | 4957 | 02371 | 4173 | 1714 | 8764 | 04841 |
| Max Sharpe | 0.89653 8492 | 9.1649E -08 | -0.2539 82648 | 0.06113 0116 | 0.05474 995 | 0.13342 6556 | 0.21678 9087 | -0.0102 69732 | 0.13736 6216 | -0.1480 45568 | -0.0877 02561 | 0.03695 3655 | 0.10819 9339 | 0.32175 5543 |

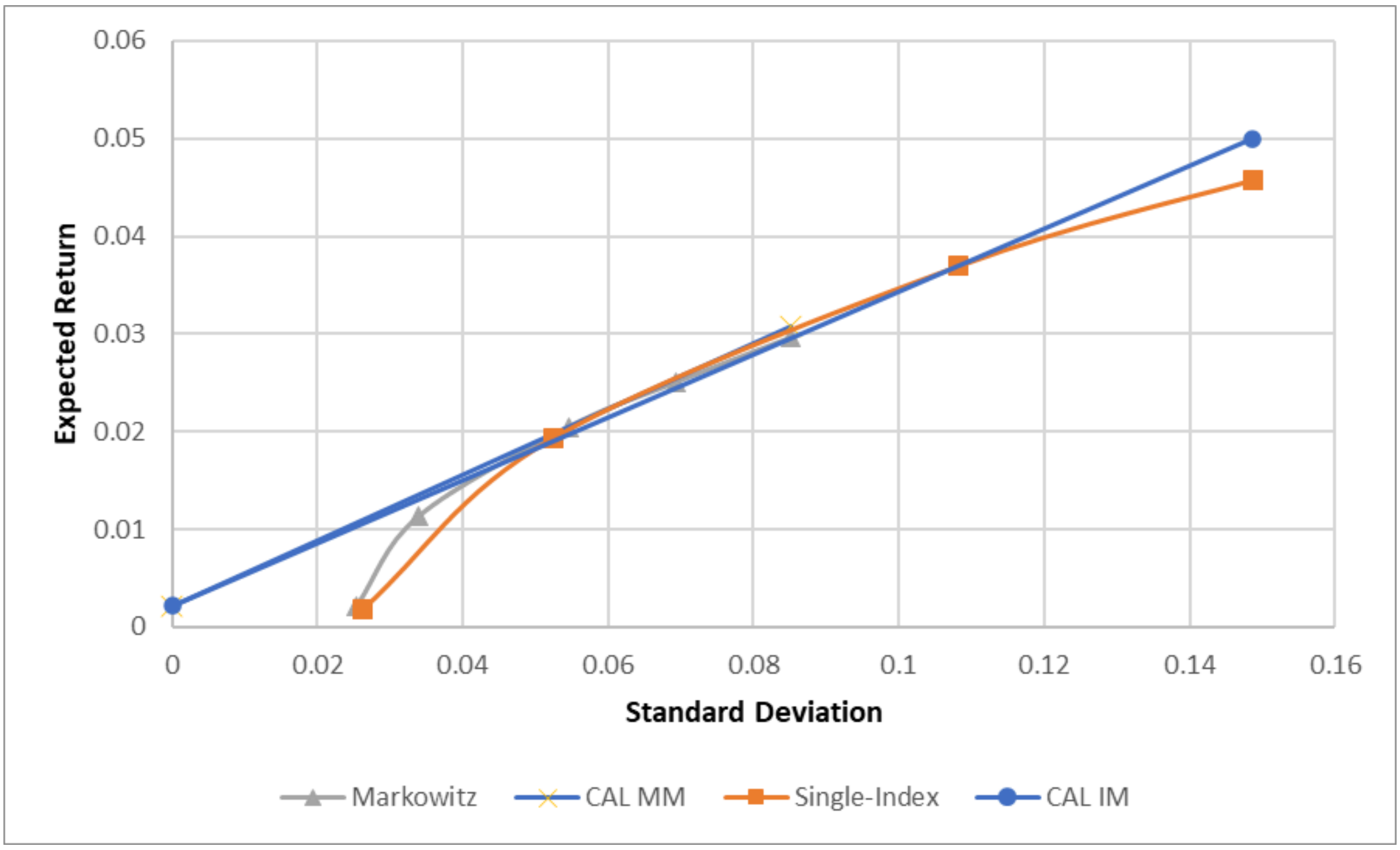

**Figure 11**

Under Constraint 1, at lower standard deviations, the Markowitz model consistently positions slightly above the Single Index model's efficient frontier. However, as standard deviation increases, the Single Index model outperforms, presenting a more favourable risk-return trade-off in higher-risk scenarios. Additionally, when examining the capital allocation line, it becomes apparent that the Markowitz model's line is slightly steeper than that of the Single Index model. This difference is due to the Markowitz model's capacity to achieve a higher maximum Sharpe ratio. The range of point dispersion of the portfolio is almost the same in both models.

**Constraint 2**

**Table 8**: Minimum variance portfolio and maximum sharpe portfolio in MM

| | INAR | CELC | AXIA | HLCB | HLBB | IHHH | HTHB | GENM | DIAL | GENT | Market | Return | StDev | Sharpe |
|---|---|---|---|---|---|---|---|---|---|---|---|---|---|---|
| Minimum Variance | 0.002196687 | 0.113339589 | -0.131830633 | -0.003712517 | 0.201984227 | 0.224250926 | -0.019074397 | 0.016073287 | -0.000801827 | -0.127436399 | 0.725011068 | 0.002149832 | 0.025442835 | 0.000389683 |
| Max Sharpe | 0.606424303 | -0.05869914 | -0.13873228 | 0.470330121 | 0.446833707 | 0.999996013 | 0.005449002 | 0.016624551 | 0.261551392 | -0.60977767 | -0.999999999 | 0.034022365 | 0.084799413 | 0.375974856 |

**Table 9**: Minimum variance portfolio and maximum sharpe portfolio in IM

| | INAR | CELC | AXIA | HLCB | HLBB | IHHH | HTHB | GENM | DIAL | GENT | Market | Return | StDev | Sharpe |
|---|---|---|---|---|---|---|---|---|---|---|---|---|---|---|
| Minimum Variance | 0.002373492 | 0.052599037 | -0.103098433 | 0.052044736 | 0.155852066 | 0.192401451 | 0.004248951 | -0.038791419 | -0.030327122 | -0.072005525 | 0.784702765 | 0.001731634 | 0.026158764 | -0.015607917 |
| Max Sharpe | 0.878718376 | -0.20740668 | -0.774071288 | 0.309724531 | 0.290045756 | 0.999971419 | 0.21411292 | -0.021389529 | 0.528893739 | -0.218599244 | -0.999999999 | 0.048821438 | 0.124381005 | 0.375310688 |

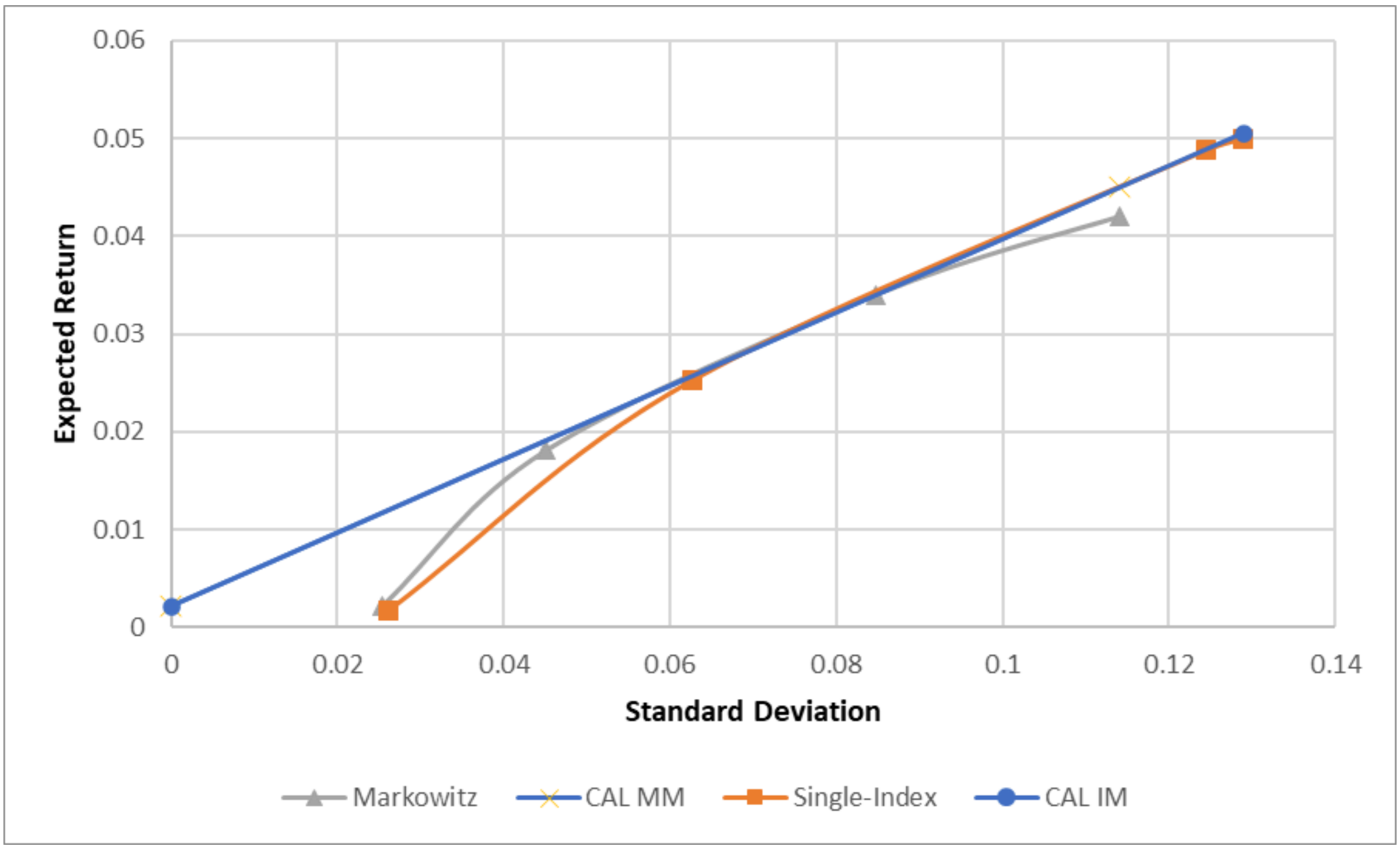

**Figure 12**

Under Constraint 2, again, the comparison of efficient frontiers between the Markowitz and Single Index models reveals consistent patterns. At lower standard deviations, the Markowitz model consistently positions slightly above the Single Index model's efficient frontier. However, as standard deviation increases, the Single Index model outperforms, presenting a more favourable risk-return trade-off in higher-risk scenarios. Interestingly, the capital allocation lines for both models are roughly the same, due to their very similar maximum Sharpe ratios. The range of point dispersion of the portfolio is almost the same in both models.

**Constraint 3**

**Table 10**: Minimum variance portfolio and maximum sharpe portfolio in MM

| | INAR | CELC | AXIA | HLCB | HLBB | IHHH | HTHB | GENM | DIAL | GENT | Market | Return | StDev | Sharpe |
|---|---|---|---|---|---|---|---|---|---|---|---|---|---|---|
| Minimum Variance | 0.002192719 | 0.113491573 | -0.131888659 | -0.00372364 | 0.201990048 | 0.224155746 | -0.019076893 | 0.016076127 | -0.000807211 | -0.127410013 | 0.725000 | 0.002149317 | 0.025442836 | 0.000369427 |
| Max Sharpe | 11.43214108 | 3.411554267 | 0.803156204 | 8.487087625 | 19.78625986 | 17.83807239 | 2.199026468 | 3.991607809 | 10.55607593 | -6.635944759 | -70.86903686 | 0.74810589 | 1.746113937 | 0.427214947 |

**Table 11**: Minimum variance portfolio and maximum sharpe portfolio in IM

| | INAR | CELC | AXIA | HLCB | HLBB | IHHH | HTHB | GENM | DIAL | GENT | Market | Return | StDev | Sharpe |
|---|---|---|---|---|---|---|---|---|---|---|---|---|---|---|
| Minimum Variance | 0.002373292 | 0.052599382 | -0.103097903 | 0.05204525 | 0.155853576 | 0.192399546 | 0.004248873 | -0.038791442 | -0.03032638 | -0.072007074 | 0.784703 | 0.001731629 | 0.026158764 | -0.015608123 |
| Max Sharpe | 13.66581122 | 0.401303613 | -4.824068355 | 13.31936961 | 16.95002535 | 17.11569316 | 3.680729248 | 3.34525265 | 10.38778887 | -0.929611795 | -72.11229357 | 0.862552342 | 2.060363327 | 0.417602281 |

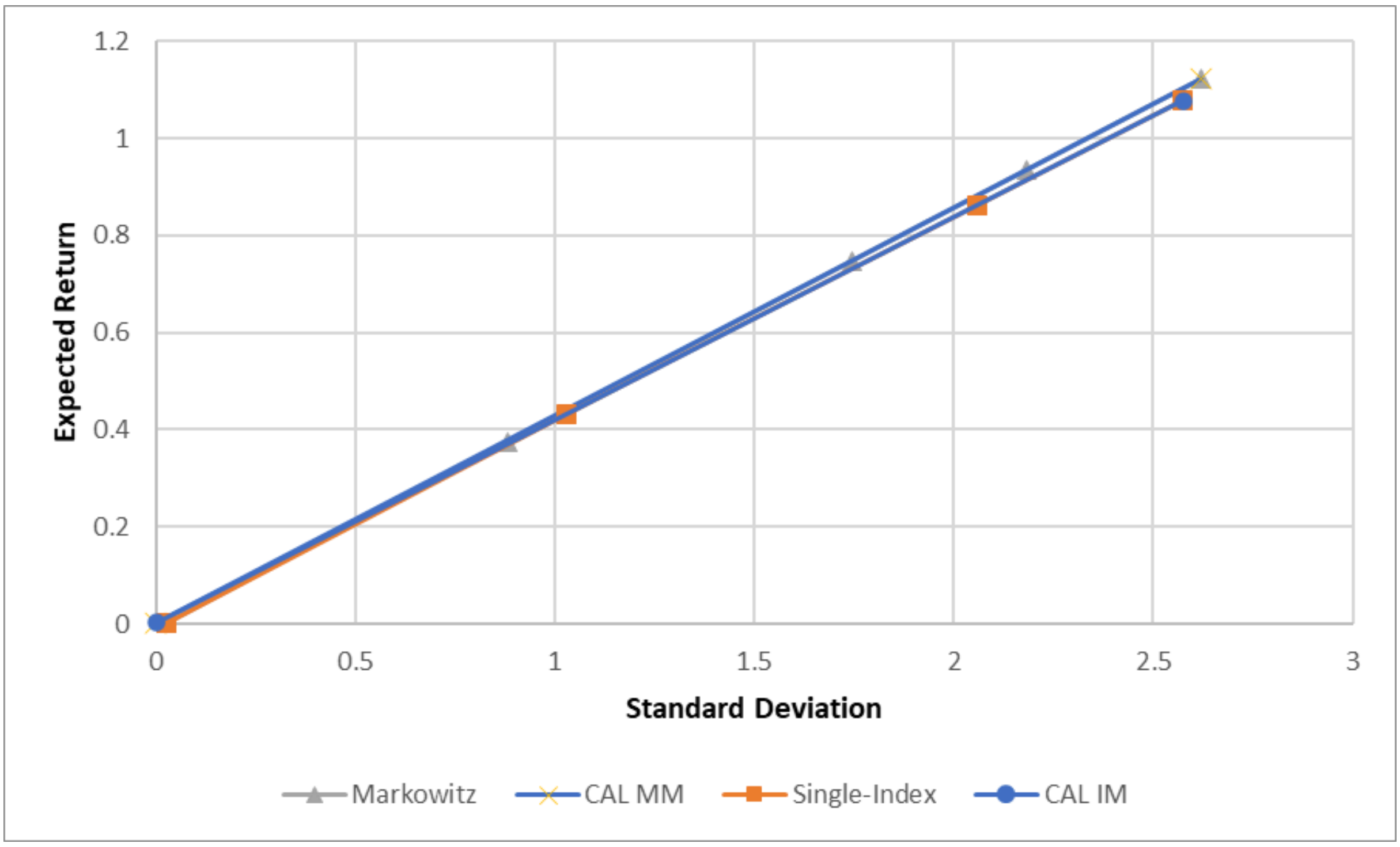

**Figure 13**

Under Constraint 3, both the Markowitz and Single Index models reveal an overlapping and straight alignment of the efficient frontier and capital allocation lines. This alignment indicates that, given the specified constraint, the portfolios generated by both models share identical risk-return characteristics. While the efficient frontier and capital allocation line for the Markowitz model are slightly steeper than those of the Single Index model, the difference is not considered significant. This marginal discrepancy suggests that, in terms of risk-adjusted returns, the Markowitz model holds a slight advantage. However, the impact of Constraint 3 minimises the distinction between the two models' portfolio optimization outcomes. The range of point dispersion of the portfolio is almost the same in both models.

**Constraint 4**

**Table 12**: Minimum variance portfolio and maximum sharpe portfolio in MM

| | INAR | CELC | AXIA | HLCB | HLBB | IHHH | HTHB | GENM | DIAL | GENT | Market | Return | StDev | Sharpe |
|---|---|---|---|---|---|---|---|---|---|---|---|---|---|---|
| Minimum Variance | 0.00159 8013 | 0.05533 5532 | 0 | 0.05004 6364 | 0.01345 2324 | 0.15912 0679 | 0.01311 0543 | 0 | 0 | 0 | 0.70733 6551 | 0.00194 9101 | 0.02734 4222 | -0.0069 78311 |
| Max Sharpe | 0.616621593 | 0 | 0 | 0 | 0 | 0.302238956 | 0.041840475 | 0 | 0.039298976 | 0 | 0 | 0.023426082 | 0.075208973 | 0.283026925 |

**Table 13**: Minimum variance portfolio and maximum sharpe portfolio in IM

| | INAR | CELC | AXIA | HLCB | HLBB | IHHH | HTHB | GENM | DIAL | GENT | Market | Return | StDev | Sharpe |
|---|---|---|---|---|---|---|---|---|---|---|---|---|---|---|
| Minimum Variance | 0.002621802 | 0.058107064 | 0 | 0.057494736 | 0.172173352 | 0.212547021 | 0.004693776 | 0 | 0 | 0 | 0.49236225 | 0.001689937 | 0.027494304 | -0.016366318 |
| Max Sharpe | 0.589263749 | 0 | 0 | 0 | 0 | 0.241950551 | 0.107491675 | 0 | 0.061294025 | 0 | 0 | 0.023093814 | 0.072980149 | 0.287117751 |

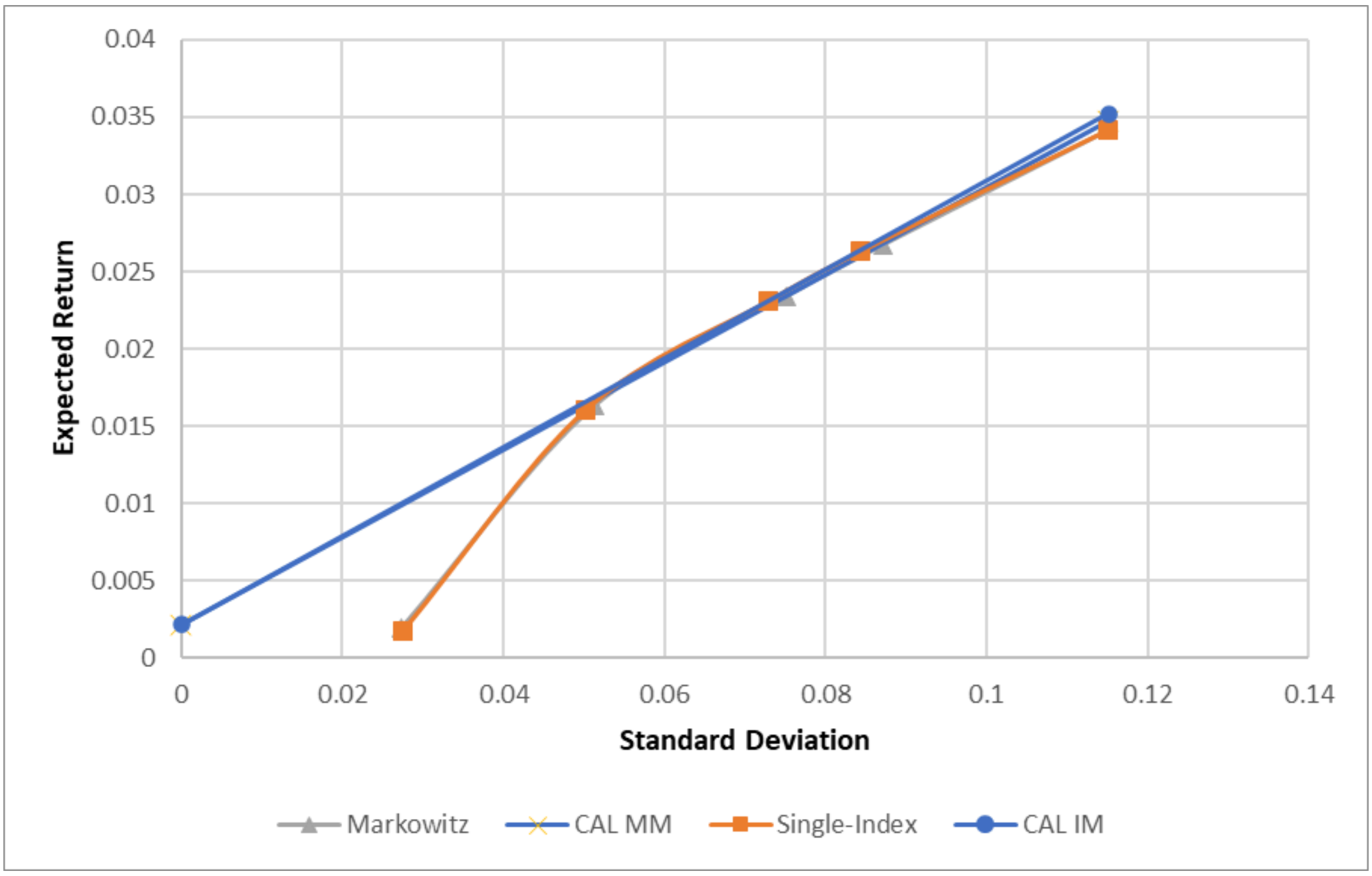

**Figure 14**

Under Constraint 4, the efficient frontiers of both the Markowitz and Single Index models are nearly identical, indicating similar risk-return trade-offs in portfolio optimization. Likewise, the capital allocation lines for both models closely align, with the distinction that the capital allocation line for the Single Index model is slightly steeper. This slight difference is attributed to the Single Index model achieving a slightly higher maximum Sharpe ratio. The range of point dispersion of the portfolio is almost the same in both models.

**Constraint 5**

**Table 14**: Minimum variance portfolio and maximum sharpe portfolio in MM

| | INAR | CELC | AXIA | HLCB | HLBB | IHHH | HTHB | GENM | DIAL | GENT | Market | Return | StDev | Sharpe |
|---|---|---|---|---|---|---|---|---|---|---|---|---|---|---|
| Minimum Variance | 0.026950642 | 0.000233139 | 0.010067645 | 0.044331327 | 0.45277249 | 0.398968882 | 0.025379595 | 0.036520695 | 0.104083987 | -0.0993084 | 0 | 0.005952205 | 0.029379846 | 0.129758583 |
| Max Sharpe | 0.572927326 | -5.88178E-05 | -0.435522624 | 0.312355886 | 0.348754874 | 0.619104506 | -0.068660687 | 0.001484014 | 0.205580141 | -0.555964605 | 0 | 0.029336065 | 0.078926044 | 0.344577599 |

**Table 15**: Minimum variance portfolio and maximum sharpe portfolio in IM

| | INAR | CELC | AXIA | HLCB | HLBB | IHHH | HTHB | GENM | DIAL | GENT | Market | Return | StDev | Sharpe |
|---|---|---|---|---|---|---|---|---|---|---|---|---|---|---|
| Minimum Variance | 0.017797205 | 0.160705936 | -0.039168388 | 0.174342372 | 0.335187684 | 0.313847653 | 0.015790008 | 0.010804398 | 0.025410572 | -0.014717428 | 0 | 0.004626101 | 0.028944171 | 0.08589582 |
| Max Sharpe | 0.854613639 | -0.505981275 | -0.769348232 | 0.312594899 | 0.341697295 | 0.69535758 | 0.19398019 | -0.068351113 | 0.383953851 | -0.438516813 | 0 | 0.045137194 | 0.120475308 | 0.356897005 |

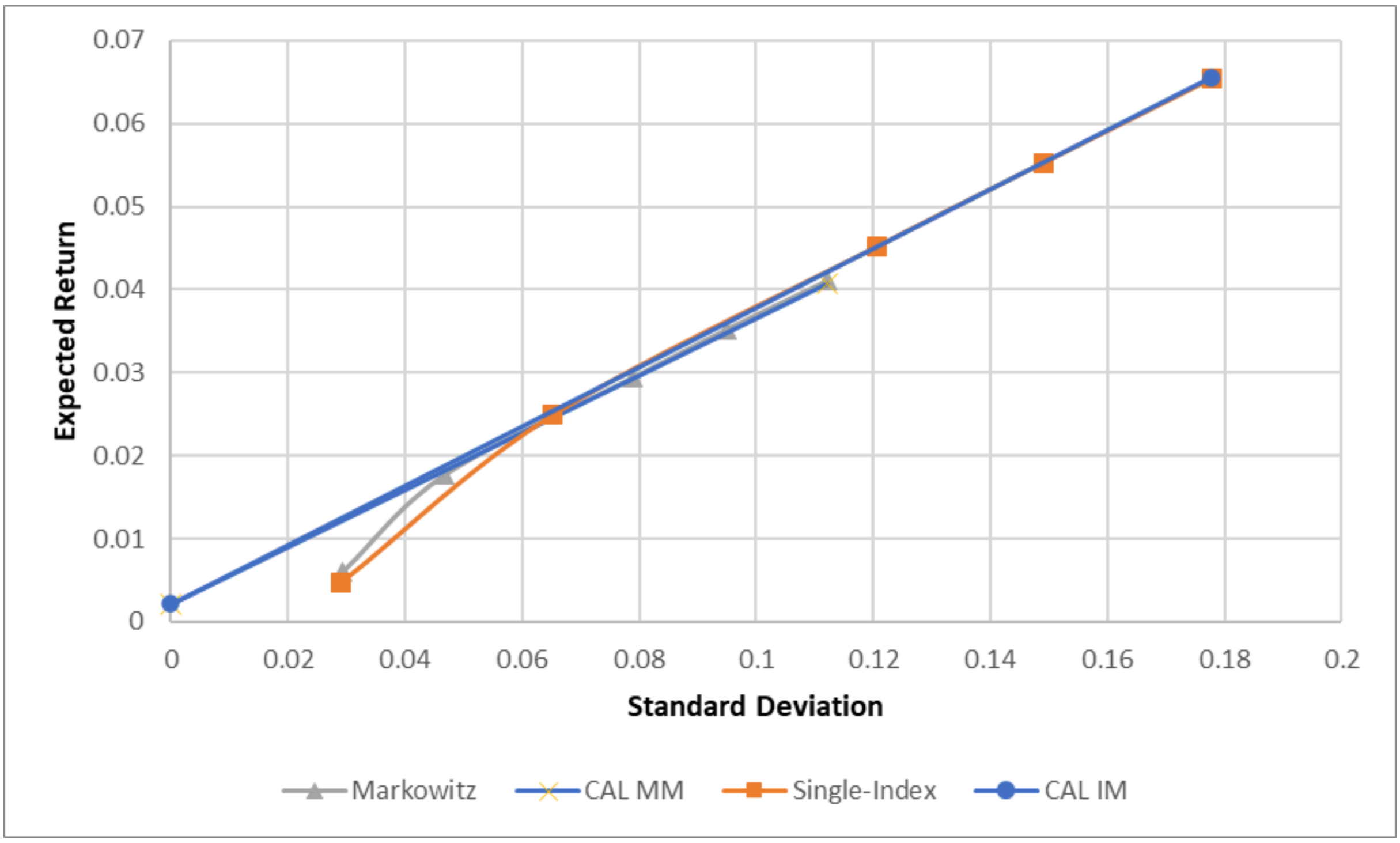

**Figure 15**

Under Constraint 5, at lower standard deviations, the Markowitz model consistently positions slightly higher than the Single Index model, indicating potential advantages in risk-adjusted returns. However, as standard deviation increases, the Single Index model's efficient frontier edges slightly above that of the Markowitz model. Despite these variations, the capital allocation lines for both models remain remarkably similar, suggesting a convergence in optimal portfolio allocations. The range of point dispersion of the portfolio is almost the same in both models.

Furthermore, there is a difference in the processing complexity between the two models. In our case, N is equal to 11 due to 10 stocks and 1 market index. The Markowitz model requires 2N+N*(N-1)/2 estimators for N stocks, resulting in 77 estimators (Shapiro, n.d.). In contrast, the Index model only needs 3N+2 estimators for N stocks, requiring only 35 estimators to achieve similar portfolio outcomes (Eric M. Aldrich, n.d.). As evident from the comparison, the Index model requires only half the number of estimators as the Markowitz model to achieve comparable results. While the difference is not significant with 10 stocks and 1 market index, as the number of stocks increases, the gap in the number of estimators between the two models will grow exponentially. The advantage of the Index model becomes more prominent as the number of stocks increases, owing to the simplification process that reduces estimators; however, it also increases the potential inaccuracy of returns and risks (Gallego, 1999). Hence, although our research found a small difference between the two models, as we include more stocks, there might be a bigger gap in how accurate the two models are.

# 4. Conclusion

From our research, we have discovered from the analysis of the two different models that both models result in nearly identical results for the efficient frontier and capital allocation line. Thus, there is no noticeable difference between using the Markowitz Model or the Single Index Model when constructing one's portfolio. However, the Markowitz Model requires far more inputs compared to the Single Index Model, especially for more complicated portfolios with large numbers of investment candidates. Therefore, in real-world use, the Single Index Model is more practical than the Markowitz Model as it can be easily scaled up when the number of assets in our portfolio increases. Thus, our research has shown that investors should consider applying the Single Index Model to compute their optimal portfolios, although the Markowitz Model can also be used provided the number of assets in the investor's portfolio is not too large as this will lead to a large number of inputs that the investor must furnish. We hope our research can also provide a guideline to potential investors in the Malaysian stock market on an efficient portfolio that will help them minimise their investment risk.

A limitation of our research is the number of stocks is relatively low at only 10 stocks and one market index. Therefore, the comparison between the MM and IM models may not be indicative of actual real-world usage as many investors hold complex portfolios with significantly higher numbers of stocks. When the number of stocks in a portfolio increases, the differences between the two models may become more apparent and it may no longer be feasible to regard both models as having similar accuracy. In the future, we hope to perform research involving portfolios with even greater numbers of assets to provide a more conclusive solution on the differences between the two models.